\documentclass[aps,prl,twocolumn,showpacs,preprintnumbers,superscriptaddress, amsmath,amssymb,10pt]{revtex4-2}

\usepackage{graphicx}
\usepackage{dcolumn}
\usepackage{bm}
\usepackage{amsmath}
\usepackage{amssymb}
\usepackage{physics}
\usepackage{color}
\usepackage{hyperref, booktabs}
\usepackage{todonotes}
\usepackage[justification=raggedright,
            singlelinecheck=false]{caption}
\usepackage{subcaption}

\begin{document}


\title{Generalized Langevin Models of Linear Agent-Based Systems: Strategic Influence Through Environmental Coupling}

\author{Semra G\"und\"u\c{c}}
\affiliation{Computational Mathematics, Science and Engineering, Michigan State University, East Lansing, Michigan 48824, USA}
\affiliation{Computer Engineering Department, Ankara University, Ankara 06880, Turkiye}

\author{David J. Butts}
\affiliation{Information Systems \& Modeling, Los Alamos National Laboratory, Los Alamos, New Mexico 87545, USA}

\author{Michael S. Murillo}
\affiliation{Computational Mathematics, Science and Engineering, Michigan State University, East Lansing, Michigan 48824, USA}

\date{\today}

\begin{abstract}
Agent-based models typically treat systems in isolation, discarding environmental coupling as either computationally prohibitive or dynamically irrelevant. We demonstrate that this neglect misses essential physics: environmental degrees of freedom create memory effects that fundamentally alter system dynamics. By systematically transforming linear update rules into exact generalized Langevin equations, we show that unobserved environmental agents manifest as memory kernels whose timescales and coupling strengths are determined by the environmental interaction spectrum. Network topology shapes this memory structure in distinct ways: small-world rewiring drives dynamics toward a single dominant relaxation mode, while fragmented environments sustain multiple persistent modes corresponding to isolated subpopulations. We apply this framework to covert influence operations where adversaries manipulate target populations exclusively via environmental intermediaries. The steady-state response admits a random-walk interpretation through hitting probabilities, revealing how zealot opinions diffuse through the environment to shift system agent opinions toward the zealot mean—even when zealots never directly contact targets.
\end{abstract}

\maketitle

Complex systems rarely evolve in isolation. Agent-based models~\cite{bonabeau2002agent,epstein2006generative} must therefore confront a fundamental choice which is to include environmental coupling at potentially prohibitive computational cost, or neglect it and risk missing essential dynamics. The COVID-19 pandemic starkly illustrated this trade-off. Models incorporating travel networks and regional coupling proved indispensable for policy decisions~\cite{anderson2020,chinazzi2020,wells2020}, yet their computational demands remain prohibitive for many applications. This challenge extends beyond epidemiology to social influence campaigns, economic contagion, and ecological systems, where environmental effects dominate behavior but full environmental simulation remains intractable.  We demonstrate that for linear agent-based systems, this computational dilemma admits an elegant resolution. Environmental coupling naturally manifests as memory kernels and fluctuating forces that encode the cumulative effect of unobserved degrees of freedom, thereby capturing essential environmental information without explicit simulation. 

Linear models deserve special attention for three reasons. First, many complex systems admit linearization near equilibrium points or consensus states, where perturbative dynamics determine stability and response \cite{vandendriessche2002reproduction}. Second, data-driven methods like Dynamic Mode Decomposition naturally produce linear models that capture dominant system behavior~\cite{schmid2022dynamic,haller2024datadriven}. Third, several foundational models across diverse domains, DeGroot opinion dynamics in social systems \cite{degroot1974}, Leslie matrices for age-structured populations \cite{leslie1945use,caswell2001matrix}, and Leontief input-output models in economics \cite{leontief1986input,miller2009input}, are inherently linear. Far from being restrictive, linearity provides the tractability needed to derive exact solutions for environmental coupling while capturing essential dynamics in regimes where nonlinear effects remain perturbative. Most importantly, linear systems offer unparalleled analytical transparency, making them the natural starting point for understanding how environmental coupling shapes collective behavior.

We first derive exact generalized Langevin equations \cite{zwanzig2001nonequilibrium,vanKampen2007}  for partitioned agent systems, revealing how environmental coupling manifests as memory kernels and fluctuating forces. We then analyze DeGroot opinion dynamics on synthetic and real networks, introducing operator centrality, a framework quantifying agent influence and vulnerability through spectral properties of memory operators. Finally, we demonstrate strategic implications through covert influence operations where adversaries manipulate targets exclusively via environmental intermediaries, deriving optimal zealot placement strategies and validating predictions on empirical social networks.

The mathematical foundation for our approach draws inspiration from the Mori-Zwanzig formalism in statistical mechanics~\cite{mori1965transport,zwanzig1961memory}, which uses projection operators to partition the total space into observed system variables and unobserved environmental degrees of freedom. However, the linear agent-based models enable a more direct approach. We can partition agents into system and environment using block matrices, avoiding projection operators while achieving exact dimensionality reduction. The resulting generalized Langevin equation emerges naturally from the block structure, with memory kernels encoding how past system states influence future dynamics through environmental mediation.

Consider $N$ agents with state vector $\mathbf{x} \in \mathbb{R}^N$ evolving according to the linear discrete update rule
\begin{equation}
\mathbf{x}(n+1) = \mathbf{D}\mathbf{x}(n),
\label{eq:discrete_update}
\end{equation}
where the discrete update matrix $\mathbf{D}$ encodes agent interactions and environmental influences, and $n$ denotes the discrete time step.  To systematically derive open system dynamics, we partition all $N$ agents into $\mathbf{x}_s \in \mathbb{R}^{N_s}$ (system agents of primary interest) and $\mathbf{x}_b \in \mathbb{R}^{N_b}$ (bath or environmental agents), where $N_s + N_b = N$. The partitioned dynamics become
\begin{equation}
\begin{pmatrix}
\mathbf{x}_{s}(n+1) \\
\mathbf{x}_{b}(n+1)
\end{pmatrix} = 
\begin{pmatrix}
\mathbf{D}_{ss} & \mathbf{D}_{sb} \\
\mathbf{D}_{bs} & \mathbf{D}_{bb}
\end{pmatrix}
\begin{pmatrix}
\mathbf{x}_{s}(n) \\
\mathbf{x}_{b}(n)
\end{pmatrix},
\label{partitioned_dynamic}
\end{equation}
where $\mathbf{D}_{ss}$ governs direct interactions within the system, $\mathbf{D}_{bb}$ captures internal bath dynamics, $\mathbf{D}_{sb}$ represents bath-to-system influence, and $\mathbf{D}_{bs}$ describes system-to-bath influence. 

The bath equation, 
\begin{equation}
\mathbf{x}_b(n+1) = \mathbf{D}_{bs}\mathbf{x}_s(n) + \mathbf{D}_{bb}\mathbf{x}_b(n),
\end{equation}
can be solved exactly by iteration, which yields
\begin{equation}
\mathbf{x}_b(n) = \mathbf{D}_{bb}^n\mathbf{x}_b(0) + \sum_{k=0}^{n-1} \mathbf{D}_{bb}^{n-1-k}\mathbf{D}_{bs}\mathbf{x}_s(k).
\label{eq:bath_solution}
\end{equation}
This reveals the dual nature of environmental dynamics which are autonomous evolution from initial conditions (capturing pre-existing environmental state) and responsive evolution driven by system feedback (encoding bidirectional coupling).  Substituting Eq.~\eqref{eq:bath_solution} into the system equation yields the exact discrete-time generalized Langevin equation (GLE),
\begin{equation}
\mathbf{x}_s(n+1) = \mathbf{D}_{ss}\mathbf{x}_s(n) + \sum_{k=0}^{n-1} \mathbf{K}_d(n-k)\mathbf{x}_s(k) + \boldsymbol{\eta}_d(n),
\label{eq:discrete_gle}
\end{equation}
with memory kernel,
\begin{equation}
\mathbf{K}_d(m) = \begin{cases}
\mathbf{D}_{sb}\mathbf{D}_{bs} & \text{if } m = 0 \\
\mathbf{D}_{sb}\mathbf{D}_{bb}^{m-1}\mathbf{D}_{bs} & \text{if } m \geq 1,
\end{cases}
\label{eq:discrete_kernel}
\end{equation}
and noise,
\begin{equation}
\boldsymbol{\eta}_d(n) = \mathbf{D}_{sb}\mathbf{D}_{bb}^n\mathbf{x}_b(0).
\label{eq:discrete_noise}
\end{equation}
Note that the linearity of the dynamics allows direct algebraic elimination of bath variables — projection operators, essential for nonlinear systems in the traditional Mori-Zwanzig formalism, are not needed. Here, the noise is deterministic given bath initial conditions but appears random from the system's perspective when $\mathbf{x}_b(0)$ is unknown.

To construct the continuous-time analog, we introduce time step $\Delta t$ and assume the discrete update matrix has the expansion
\begin{equation}
\mathbf{D} = \mathbf{I} + \Delta t \cdot \mathbf{M} + O((\Delta t)^2),
\label{eq:discrete_continuous_connection}
\end{equation}
where $\mathbf{M}$ is the instantaneous rate matrix. Taking the limit $\Delta t \to 0$ while keeping $n\Delta t = t$ fixed yields
\begin{equation}
\frac{d\mathbf{x}}{dt} = \mathbf{M}\mathbf{x}.
\label{eq:continuous_dynamics}
\end{equation}
Similar steps yield the continuous-time GLE
\begin{equation}
\label{eq:cont_GLE}
\frac{d\mathbf{x}_s}{dt} \\ = \mathbf{M}_{ss}\mathbf{x}_s + \int_0^t \mathbf{K}(t-s)\mathbf{x}_s(s)ds + \boldsymbol{\eta}(t),
\end{equation}
where the memory kernel is given by
\begin{equation}
\mathbf{K}(t) = \mathbf{M}_{sb}e^{\mathbf{M}_{bb}t}\mathbf{M}_{bs}
\label{eq:continuous_kernel}
\end{equation}
and the noise is
\begin{equation}
\boldsymbol{\eta}(t) = \mathbf{M}_{sb}e^{\mathbf{M}_{bb}t}\mathbf{x}_b(0).
\label{eq:continuous_noise}
\end{equation}
The memory kernel $\mathbf{K}(t)$ encodes a three-stage process: system influence enters the bath ($\mathbf{M}_{bs}$), propagates through environmental dynamics ($e^{\mathbf{M}_{bb}t}$), and returns to affect the system ($\mathbf{M}_{sb}$). Similarly, the noise $\boldsymbol{\eta}(t)$ captures how initial bath conditions, ($\mathbf{x}_b(0)$), propagate to influence the system ($\mathbf{M}_{sb}$) through the same environmental dynamics ($e^{\mathbf{M}_{bb}t}$).

To quantify these pathways and identify which agents are most susceptible to environmental manipulation, we introduce \textit{operator centrality}—a framework for analyzing agent importance in dynamical systems with multiple influence operators. For any influence operator $\mathbf{A} \in \mathbb{R}^{N_s \times N_s}$ arising from the GLE, such as Eq.~\eqref{eq:continuous_kernel} or Eq.~\eqref{eq:continuous_noise}, operator centrality quantifies how agents participate in influence pathways through two complementary metrics: influence (capacity to drive dynamics) and vulnerability (susceptibility to external forcing). We perform the singular value decomposition
\begin{equation}
\mathbf{A} = \sum_{k=1}^{r} \sigma_k \mathbf{u}_k \mathbf{v}_k^T,
\end{equation}
where $\sigma_1 \geq \sigma_2 \geq \cdots \geq \sigma_r > 0$ are the singular values, $\mathbf{u}_k \in \mathbb{R}^{N_s}$ are the left singular vectors, $\mathbf{v}_k \in \mathbb{R}^{N_s}$ are the right singular vectors, and $r = \text{rank}(\mathbf{A})$ is the number of non-zero singular values. This spectral decomposition reveals how agent participation is distributed across dynamical modes. We define the influence and vulnerability operator centralities of agent $i$ as:
\begin{align}
I_i^{(\mathbf{A})} &= \sum_{k=1}^r \sigma_k^2 |v_{k,i}|^2,
\label{eq:influence} \\
V_i^{(\mathbf{A})} &= \sum_{k=1}^r \sigma_k^2 |u_{k,i}|^2,
\label{eq:vulnerability}
\end{align}
where $v_{k,i}$ and $u_{k,i}$ denote the $i$-th components of the right and left singular vectors $\mathbf{v}_k$ and $\mathbf{u}_k$, respectively. The influence and vulnerability are equivalently the squared column norm $\|\mathbf{A}_{:,i}\|^2$ and squared row norm $\|\mathbf{A}_{i,:}\|^2$, respectively, capturing how input $i$ affects all outputs and how output $i$ responds to all inputs.

Crucially, influence and vulnerability need not correlate. An agent can strongly drive environmental dynamics while remaining insulated from feedback, or vice versa. This input-output asymmetry, absent in traditional centrality measures, arises from non-normal memory kernel structure and reveals vulnerabilities invisible to closed-system analysis. We now demonstrate these principles through DeGroot opinion dynamics, showing how network topology shapes memory structure and creates strategic vulnerabilities in real social networks.

In the DeGroot model, agents update their opinions as weighted averages of their neighbors' opinions. The row-stochastic trust matrix $\mathbf{T}$ encodes these weights, with $T_{ij}$ representing how much agent $i$ trusts agent $j$'s opinion. The discrete update rule $\mathbf{x}(n+1) = \mathbf{T}\mathbf{x}(n)$ corresponds exactly to Eq.~\eqref{eq:discrete_update} with $\mathbf{D} = \mathbf{T}$. For continuous-time analysis, we construct the rate matrix $\mathbf{M} = (\mathbf{T} - \mathbf{I})/\Delta t$, where $\Delta t$ represents the opinion update timescale. 

To understand how network topology shapes memory effects, we decompose Eq.~\eqref{eq:continuous_kernel} through its eigendecomposition, which yields the modal structure
\begin{equation}
\mathbf{K}(t) = \sum_{k=1}^{N_b} \mathbf{C}_k e^{\lambda_k t},
\label{eq:modal_decomposition}
\end{equation}
where the coefficient matrices
\begin{equation}
\mathbf{C}_k = \mathbf{M}_{sb}\mathbf{q}_k\mathbf{p}_k^T\mathbf{M}_{bs}
\label{eq:coupling_matrix}
\end{equation}
quantify how strongly each bath mode $k$ couples to the system. Here, $\mathbf{q}_k$ and $\mathbf{p}_k$ are the right and left eigenvectors of $\mathbf{M}_{bb}$, respectively, and $\lambda_k$ is the corresponding eigenvalue. Each mode contributes to memory with a characteristic timescale $\tau_k = -1/\text{Re}(\lambda_k)$ and coupling strength $\|\mathbf{C}_k\|_F$, where $\|\cdot\|_F$ denotes the Frobenius norm. 

\begin{figure*}[!t]
  \centering

  \begin{minipage}[t]{0.19\linewidth}\vspace{0pt}
    \includegraphics[width=\linewidth,height=3.2cm,keepaspectratio]{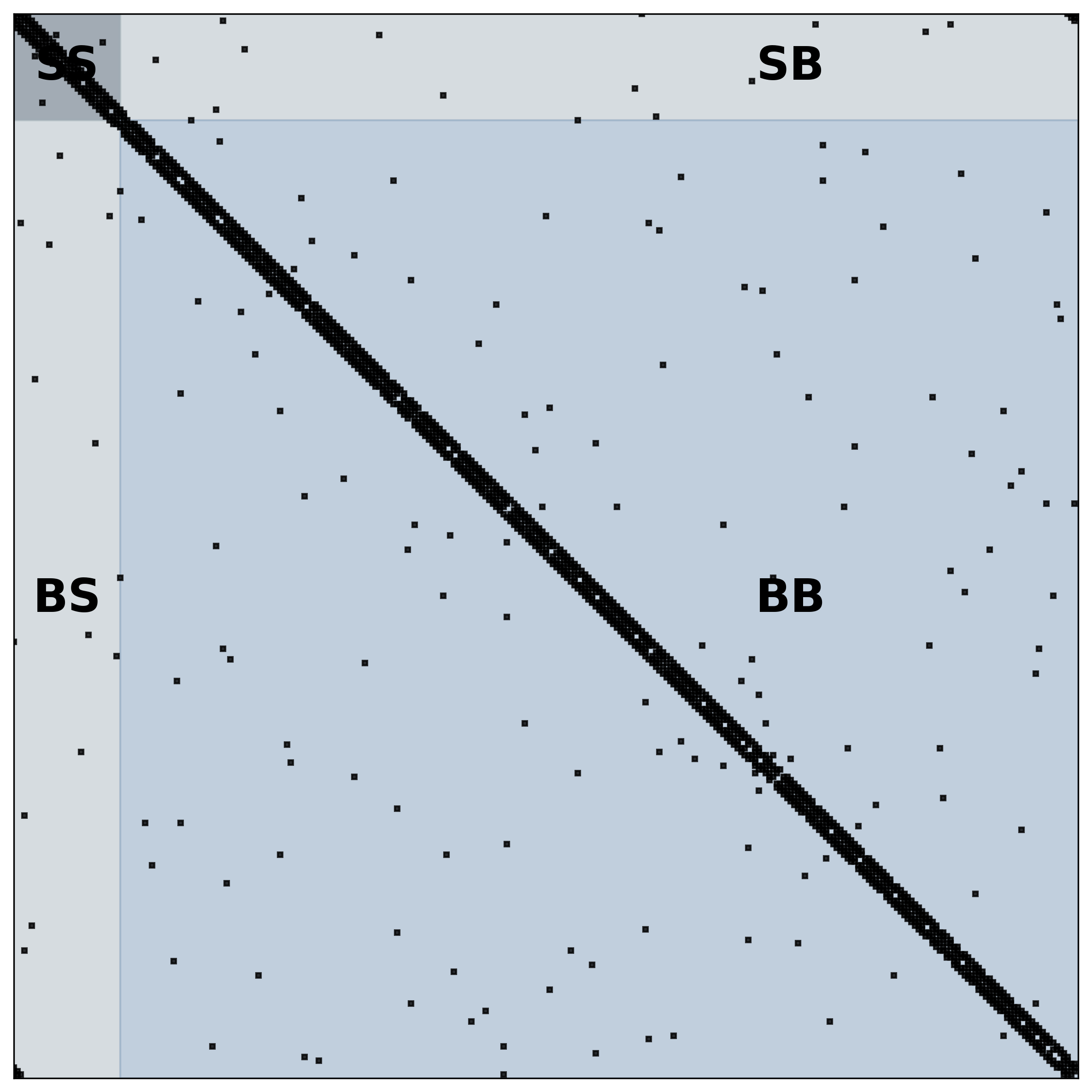}\\[2pt]
  \end{minipage}\hfill
  \begin{minipage}[t]{0.19\linewidth}\vspace{0pt}
    \includegraphics[width=\linewidth,height=3.2cm,keepaspectratio]{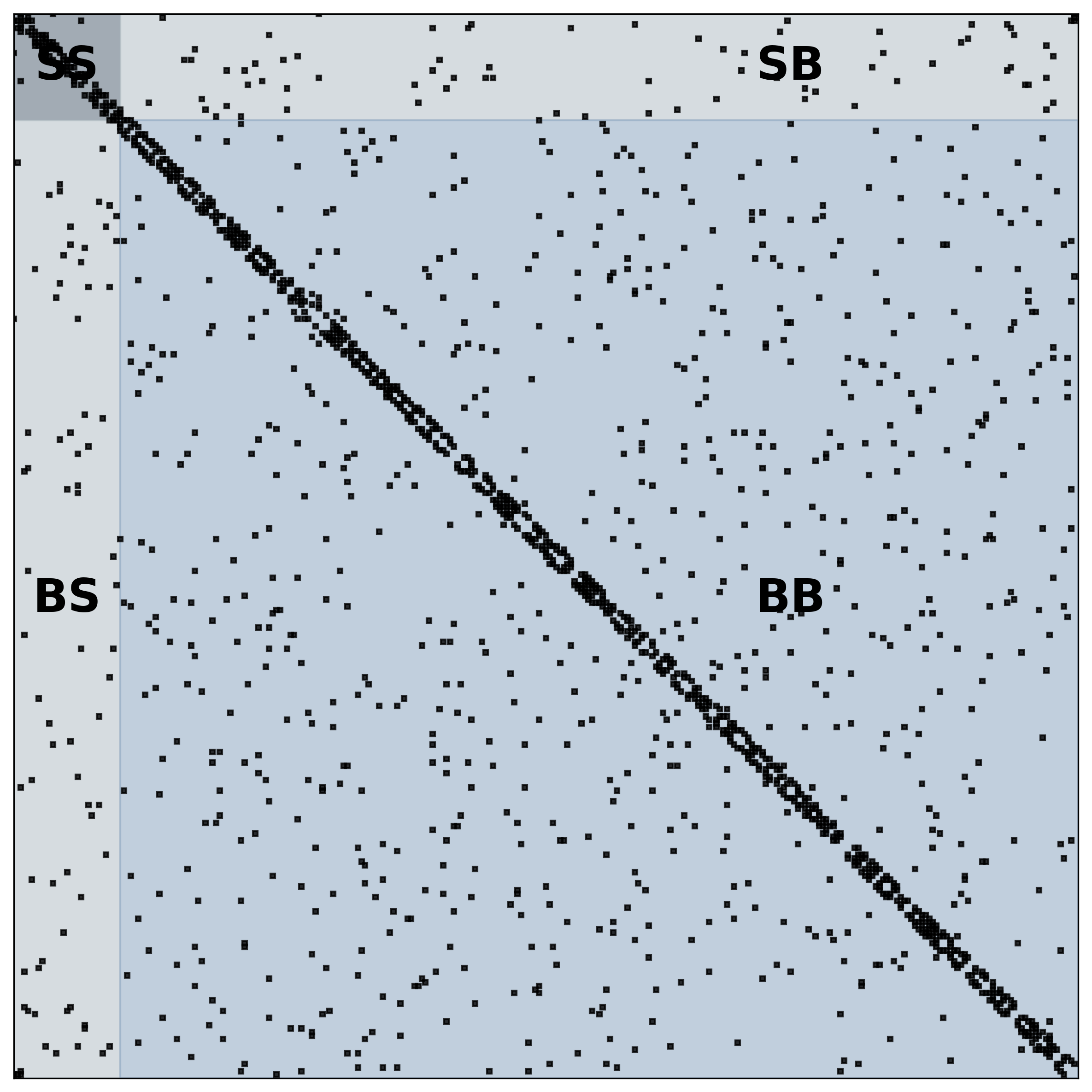}\\[2pt]
  \end{minipage}\hfill
  \begin{minipage}[t]{0.19\linewidth}\vspace{0pt}
    \includegraphics[width=\linewidth,height=3.2cm,keepaspectratio]{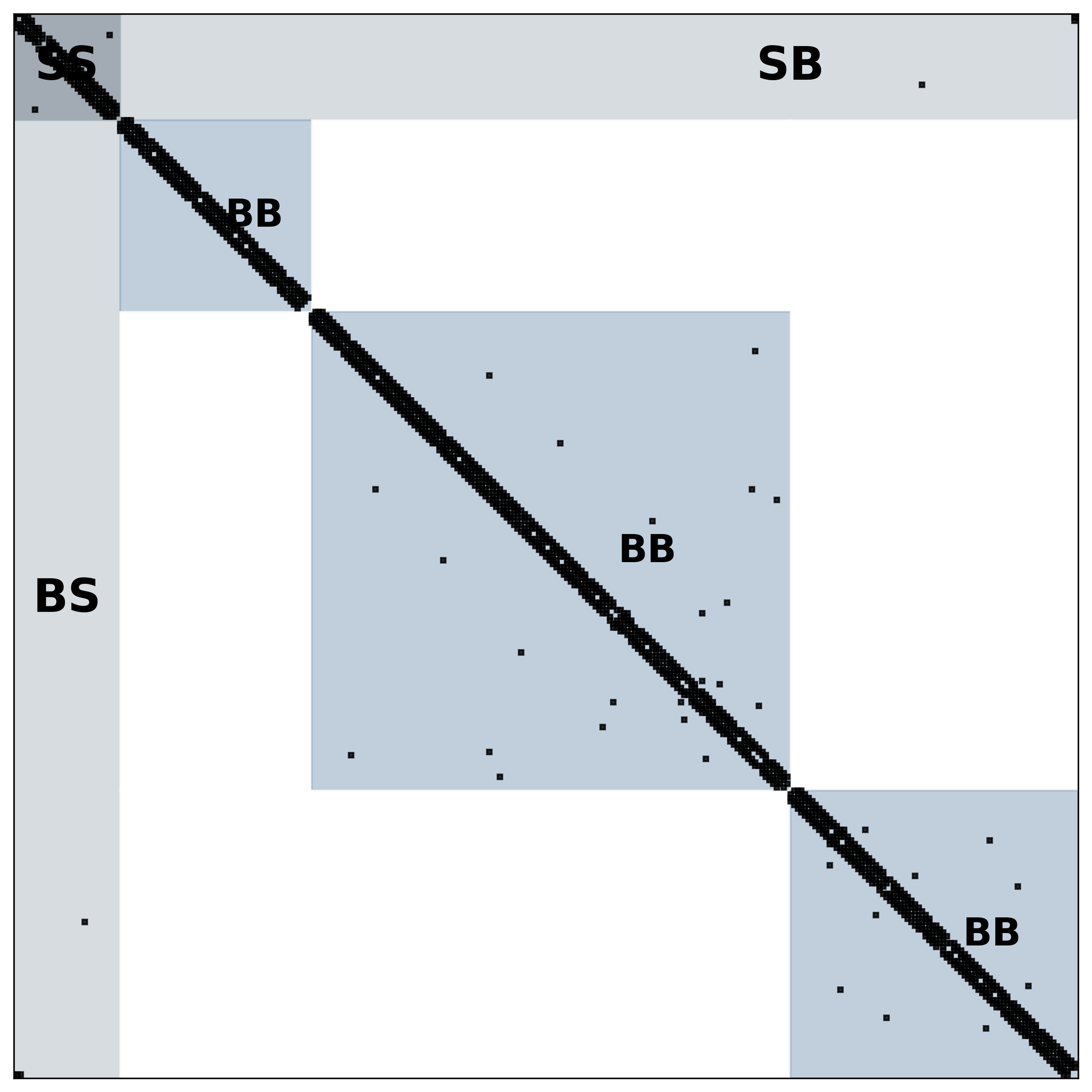}\\[2pt]
  \end{minipage}\hfill
  \begin{minipage}[t]{0.19\linewidth}\vspace{0pt}
    \includegraphics[width=\linewidth,height=3.2cm,keepaspectratio]{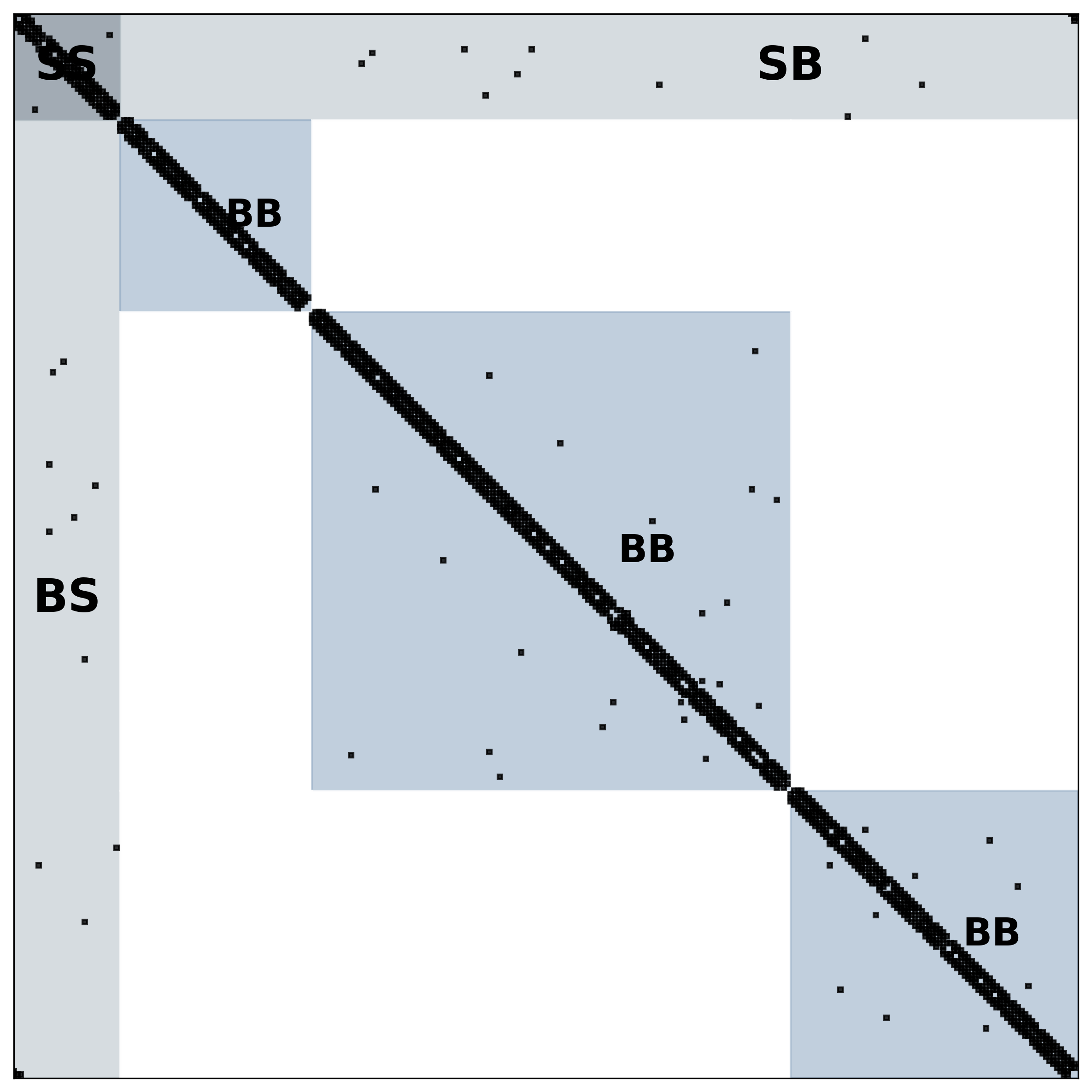}\\[2pt]
  \end{minipage}\hfill
  \begin{minipage}[t]{0.19\linewidth}\vspace{0pt}
    \includegraphics[width=\linewidth,height=3.2cm,keepaspectratio]{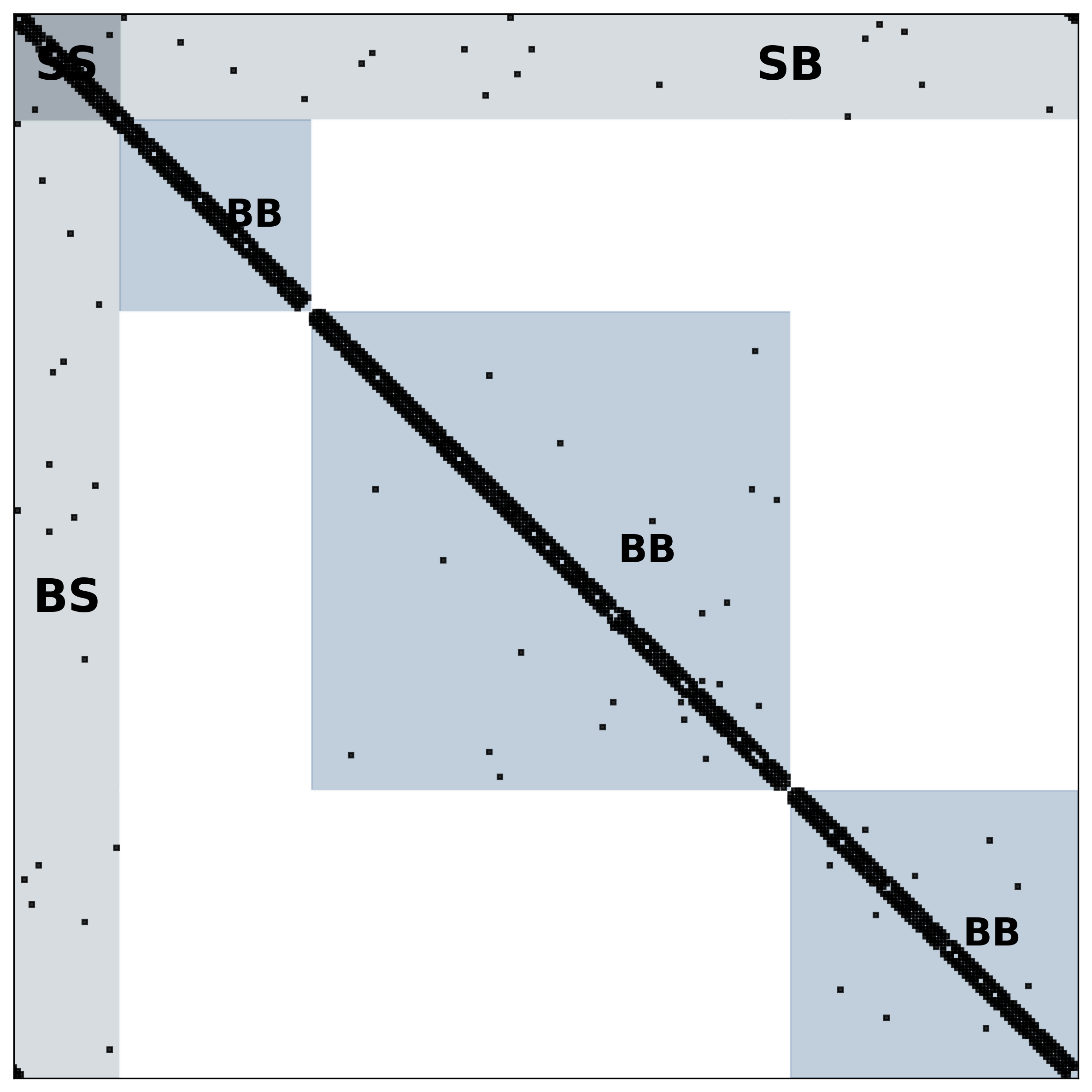}\\[2pt]
  \end{minipage}

  \vspace{0.6em}

  \begin{minipage}[t]{0.19\linewidth}\vspace{0pt}
    \includegraphics[width=\linewidth,height=3.2cm,keepaspectratio]{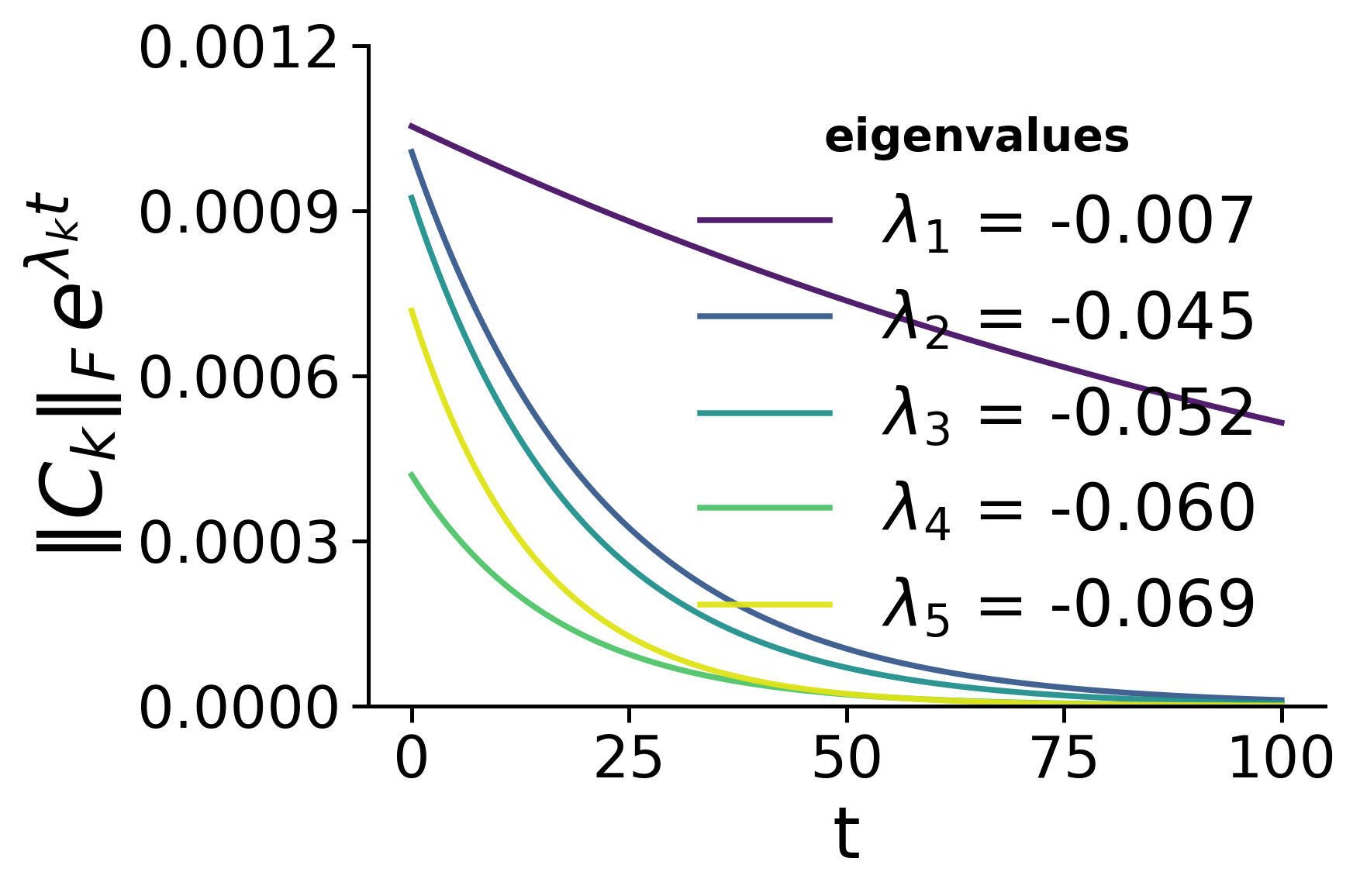}\\[2pt]
    \centering $p_{\mathrm{ws}}=0.1$
  \end{minipage}\hfill
  \begin{minipage}[t]{0.19\linewidth}\vspace{0pt}
    \includegraphics[width=\linewidth,height=3.2cm,keepaspectratio]{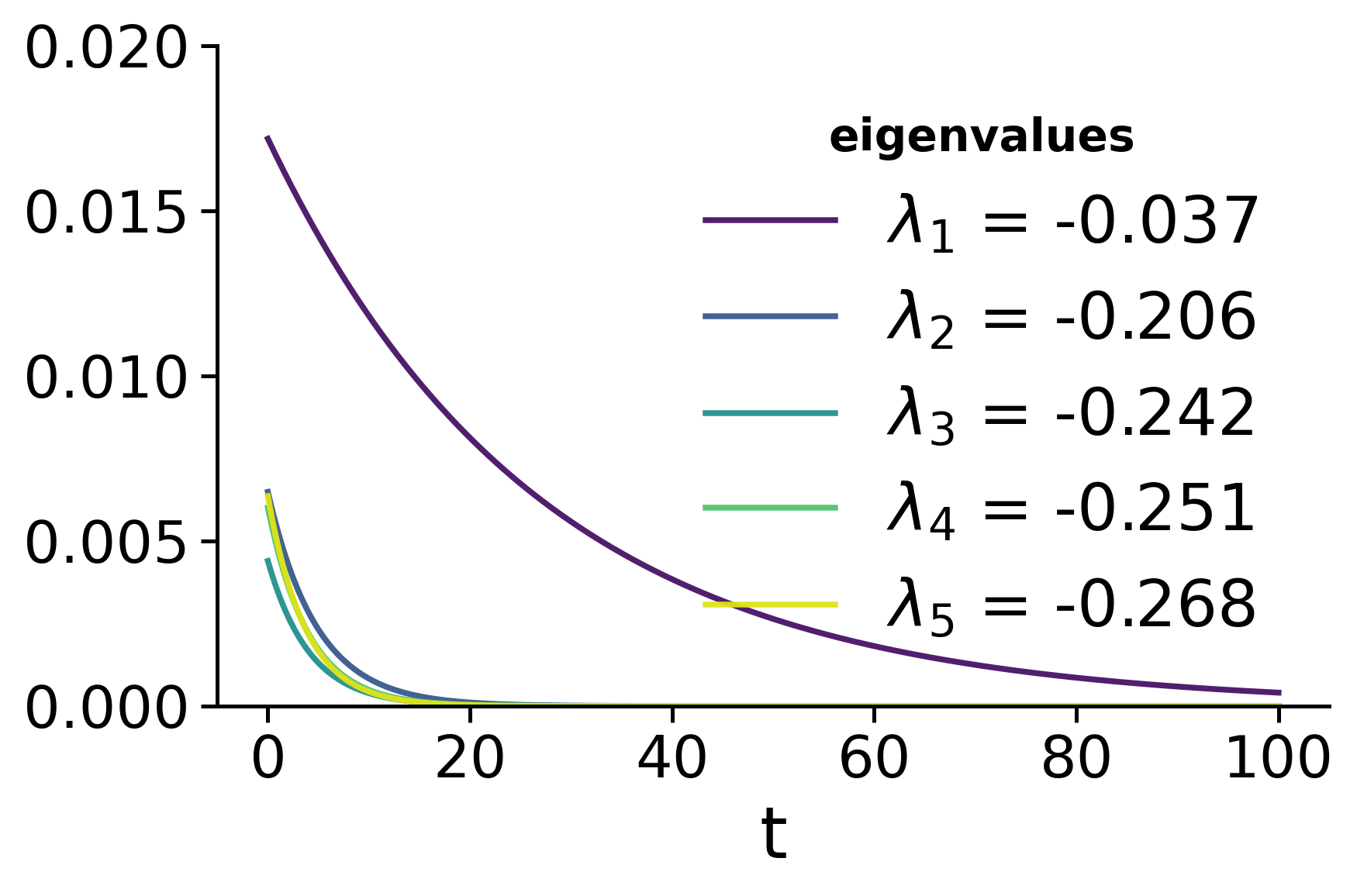}\\[2pt]
    \centering $p_{\mathrm{ws}}=0.5$
  \end{minipage}\hfill
  \begin{minipage}[t]{0.19\linewidth}\vspace{0pt}
    \includegraphics[width=\linewidth,height=3.2cm,keepaspectratio]{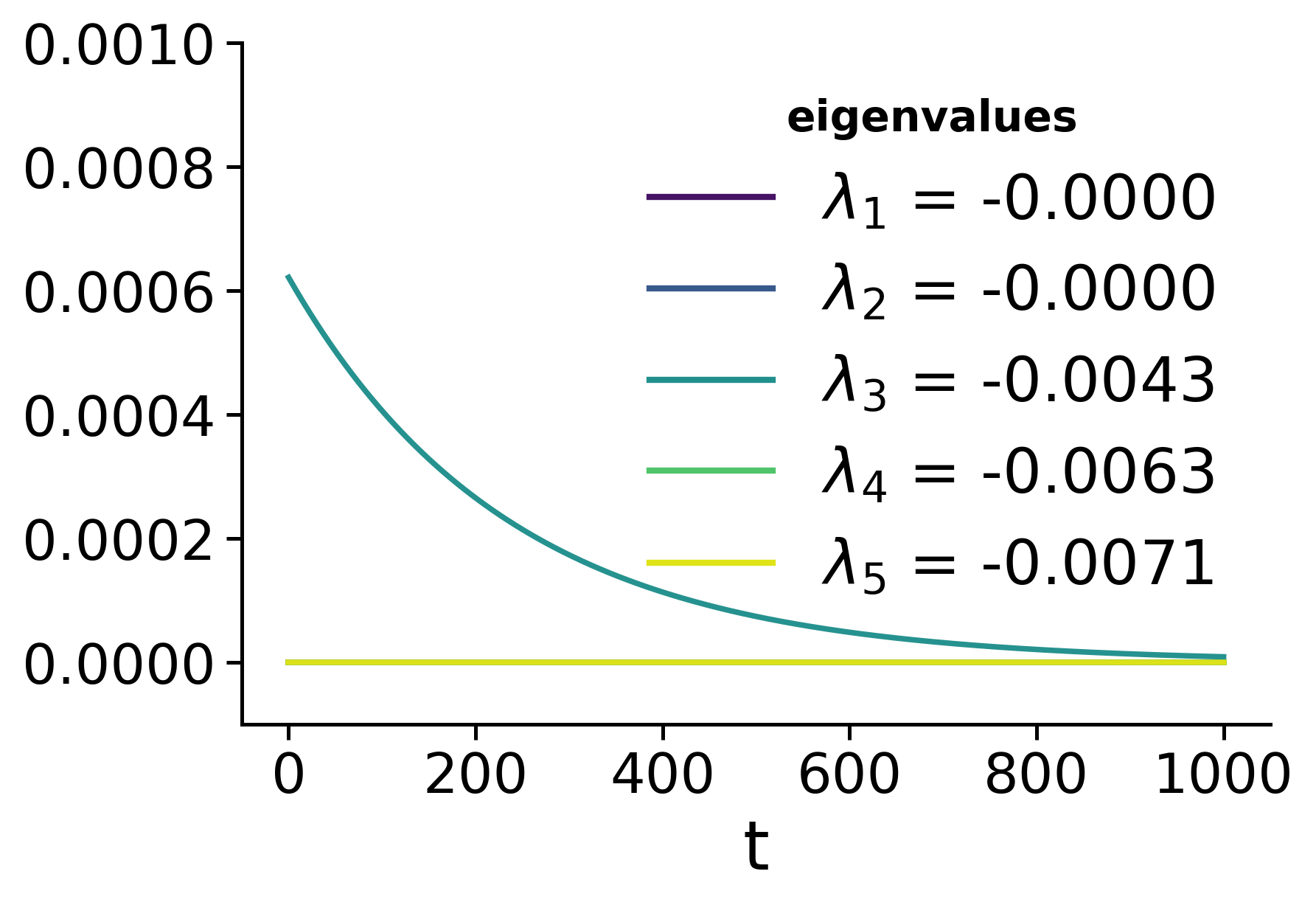}\\[2pt]
    \centering $p=0.1$
  \end{minipage}\hfill
  \begin{minipage}[t]{0.19\linewidth}\vspace{0pt}
    \includegraphics[width=\linewidth,height=3.2cm,keepaspectratio]{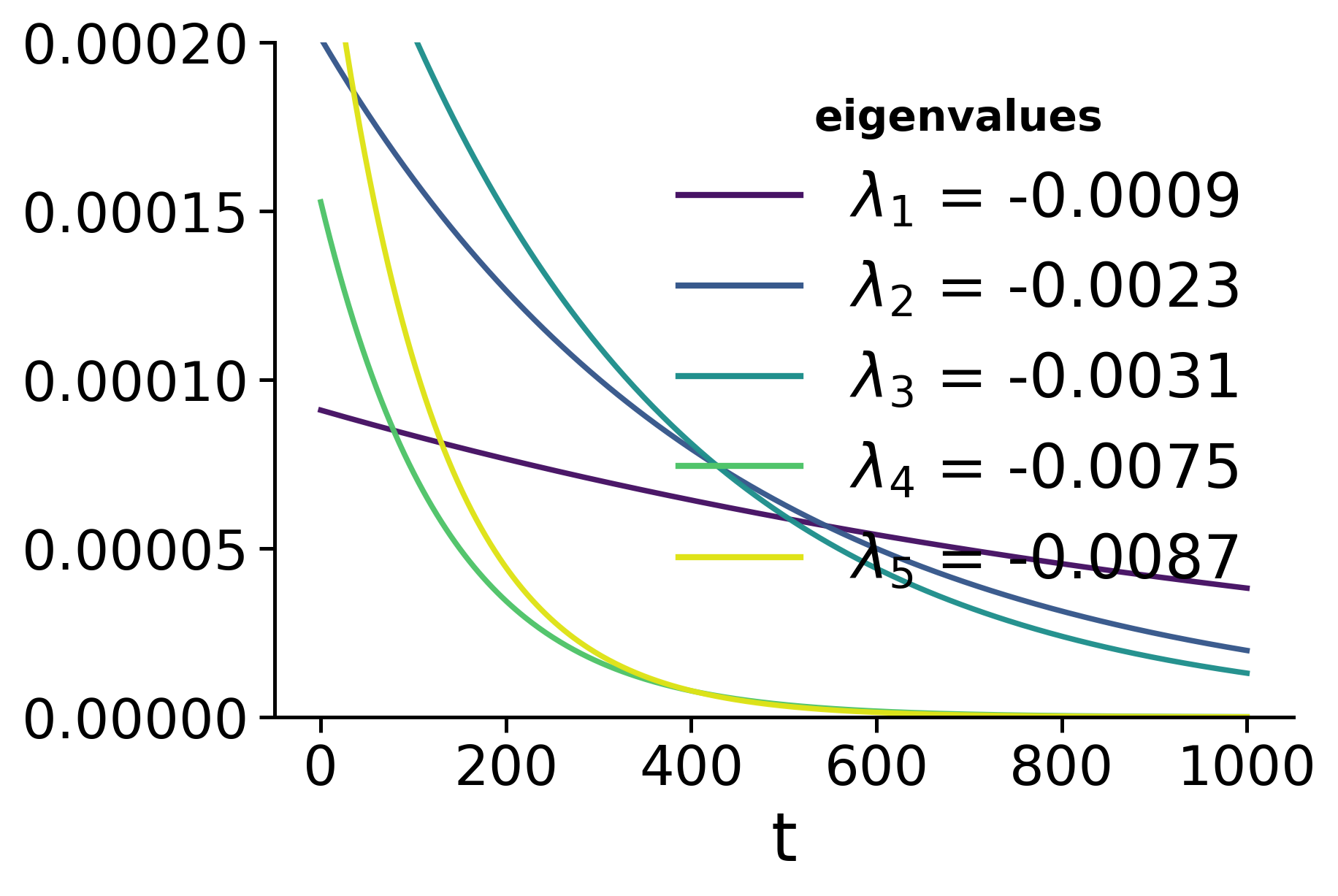}\\[2pt]
    \centering $p=0.3$
  \end{minipage}\hfill
  \begin{minipage}[t]{0.19\linewidth}\vspace{0pt}
    \includegraphics[width=\linewidth,height=3.2cm,keepaspectratio]{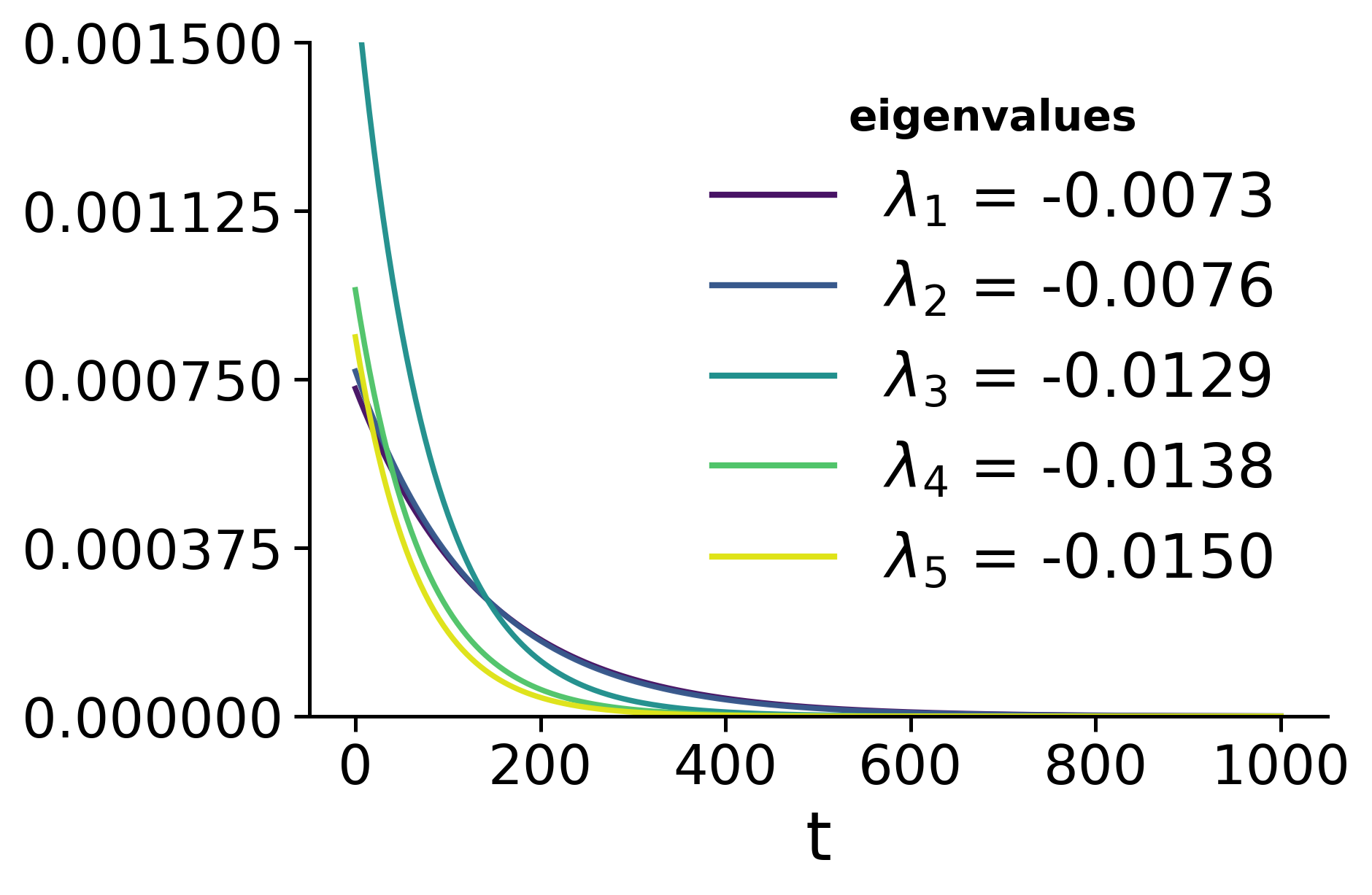}\\[2pt]
    \centering $p=1.0$
  \end{minipage}
\caption{
\textbf{Illustration of graph structure and resulting modal memory dynamics.} 
Top row shows adjacency matrices of different graph configurations. Each matrix is partitioned into system--system (upper-left, dark gray), bath--bath (lower-right, blue), and symmetric system--bath coupling blocks (off-diagonal, light gray), dots represent individual edges. Bottom row shows the corresponding modal memory dynamics. 
All graphs have $N=300$ nodes and mean degree $k=6$. 
Left two panels: Watts--Strogatz small--world graphs with rewiring probabilities $p_{\mathrm{ws}} = 0.1$ and $p_{\mathrm{ws}} = 0.5$, showing that increasing $p_{\mathrm{ws}}$ spreads edges (dots are edges) more globally, enhancing system-memory coupling and accelerating memory dynamics. 
Right three panels: echo--chamber bath configurations with fixed $p_{\mathrm{ws}} = 0.1$ and varying system-bath interaction density $f \in \{0.1, 0.3, 1.0\}$ and no interaction between bath blocks, demonstrating that higher interaction density produces stronger and more widespread memory modes.}
\label{fig:network_structure_memory}
\end{figure*}

Figure~\ref{fig:network_structure_memory} illustrates how network topology shapes memory kernel structure through two complementary variations. The first row displays adjacency matrices partitioned into system–system (upper-left, dark gray), bath–bath (lower-right, blue), and system–bath coupling blocks (off-diagonal, light gray), with dots representing individual edges. The second row shows the resulting modal memory dynamics via Eqs.~\eqref{eq:modal_decomposition}–\eqref{eq:coupling_matrix}, plotting coupling strength $\|\mathbf{C}_k\|_F$  in time. All networks contain $N=300$ nodes with mean degree $k=6$ and are partitioned into $N_s=30$ system and $N_b=270$ bath agents.

The first two columns examine Watts–Strogatz small-world graphs with rewiring probabilities $p_{\mathrm{ws}} = 0.1$ (column 1) and $p_{\mathrm{ws}} = 0.5$ (column 2). At low $p_{\mathrm{ws}}$, the adjacency matrix shows a dense diagonal band characteristic of local connectivity with sparse long-range shortcuts. As $p_{\mathrm{ws}}$ increases, rewiring distributes connections more globally, visible as increasingly dispersed off-diagonal entries. This structural transition dramatically affects memory dynamics: higher $p_{\mathrm{ws}}$ produces both stronger system–bath coupling (larger $\|\mathbf{C}_k\|_F$) and faster memory decay (more negative $\lambda_k$), reflecting enhanced information exchange through long-range pathways.

The final three columns explore bath fragmentation by partitioning the bath into three non-interacting echo chambers (visible as block-diagonal structure in $\mathbf{M}_{bb}$), while maintaining $p_{\mathrm{ws}}=0.1$ within each block. The system–bath interaction density parameter $f \in \{0.1, 0.3, 1.0\}$ controls the fraction of potential system–bath edges that are realized. At low $f$ (column 3), the off-diagonal coupling blocks are sparse; at $f=1.0$ (column 5), they are maximally dense. This variation reveals qualitatively distinct memory behavior: unlike the single-bath case where one dominant mode emerges, the echo-chamber configuration sustains multiple persistent modes—one from each isolated chamber. As $f$ increases, these modes grow stronger (larger $\|\mathbf{C}_k\|_F$) without significantly accelerating their decay, since each chamber's internal dynamics remain unchanged.

The two variations expose different aspects of memory structure. Small-world rewiring alters how the bath processes information—creating faster, more globally-coupled dynamics that amplify and dissipate memory quickly. Echo-chamber fragmentation changes what information persists—maintaining multiple independent relaxation timescales that cannot be collapsed into a single effective mode. Together, these variations demonstrate that memory kernel structure encodes both the timescales of environmental response and the multiplicity of independent influence pathways.

\begin{figure*}
     \centering
     \begin{subfigure}[t]{0.49\textwidth}
         \centering
         \includegraphics[width=\textwidth]{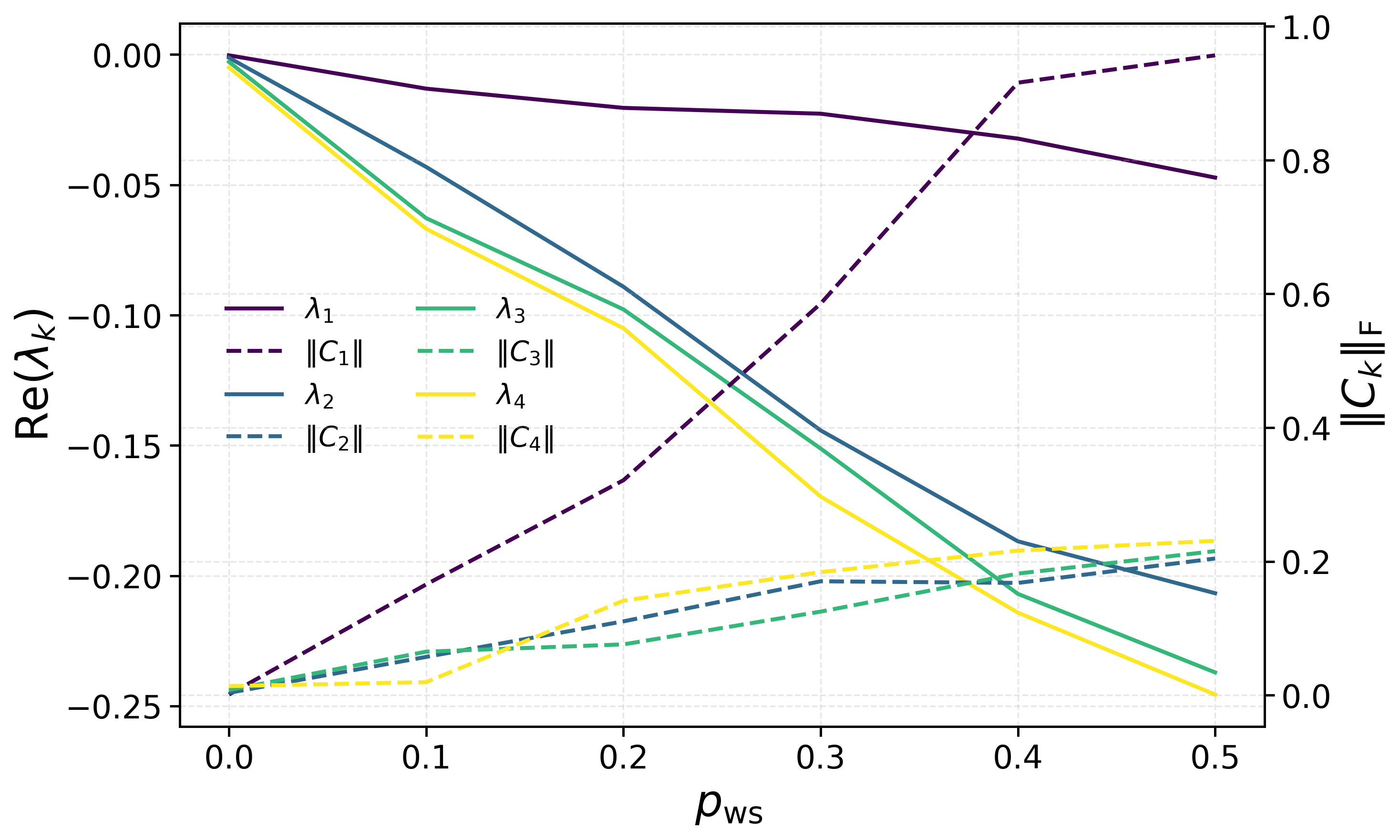}
     \end{subfigure}
     \hfill
     \begin{subfigure}[t]{0.49\textwidth}
         \centering
         \includegraphics[width=\textwidth]{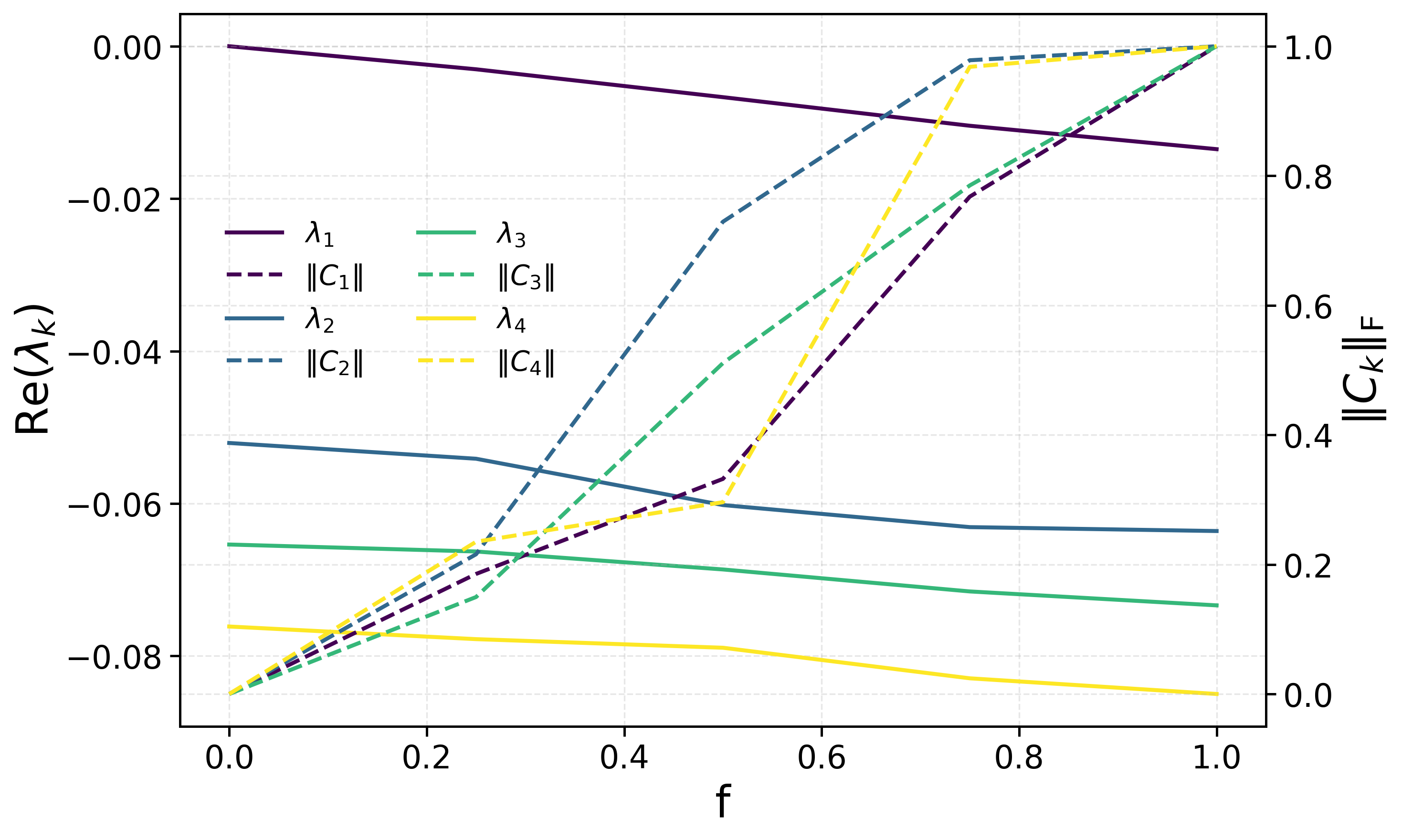}   
     \end{subfigure}
     \caption{\textbf{Eigenvalue spectra $\mathrm{Re}(\lambda_k)$ and mode coupling strengths $\|C_k\|_F$.} 
(left) Single bath system: one dominant eigenvalue governs the dynamics while the others decay. (right) Echo-chamber (block-diagonal) bath: each isolated bath block retains its own dominant eigenvalue and corresponding mode, so multiple modes remain simultaneously dominant. Note the change of scale.}
\label{fig:eig_coupling}
\end{figure*}

Figure~\ref{fig:eig_coupling} compares the eigenvalues $\lambda_k$ with their corresponding coupling strengths $\|\mathbf{C}_k\|_F$ for small--world and block--structured bath configurations. These plots illustrate how the spectral 
structure of the memory kernel evolves as either the rewiring probability $p_{\mathrm{ws}}$ of the small--world network (left panel) or the system--bath interaction density $f$ in the block configuration (right panel) is varied.

In the small--world case, when $p_{\mathrm{ws}}$ is low, the bath is nearly a regular ring. The top four 
eigenvalues of $M_{bb}$ cluster tightly near $\mathrm{Re}(\lambda_k)\approx 0$, as indicated by the solid 
lines, and their corresponding memory contributions (dashed lines) are also nearly identical. 
Because all dominant modes relax at similar rates, they provide comparable levels of coupling to 
the system. As $p_{\mathrm{ws}}$ increases and long-range edges are added, this degeneracy breaks: one 
eigenvalue remains close to zero and retains a strong memory mode, while the others move to more 
negative values and their associated coupling strengths diminish. Thus, increasing $p_{\mathrm{ws}}$ 
drives the bath toward a single dominant relaxation mode.

In the presence of echo chambers, each chamber has its own dominant eigenvalue; they are only indirectly coupled through the system. As a result, multiple eigenvalues remain near zero simultaneously, producing several long-lived memory 
modes rather than a single global relaxation mode. These block-specific modes remain spectrally 
separated due to the absence of cross-block mixing. As the system--bath interaction density increases, the coupling strengths $\|\mathbf{C}_k\|_F$ of these modes rise systematically. Thus, increasing $f$ activates and 
amplifies the memory contribution of each independent block.

Having characterized how network topology shapes memory structure, we now apply this framework to a critical security problem: covert influence operations where adversaries manipulate target populations exclusively through environmental intermediaries, never directly contacting targets. Traditional opinion control research focuses on direct manipulation or insertion of stubborn agents within the target system~\cite{jia2015opinion,acemoglu2013opinion,pasqualetti2014controllability}. Our generalized Langevin framework enables analysis of a fundamentally different scenario where manipulators must operate through the bath—capturing realistic constraints in influence campaigns where direct access is limited, detectable, or prohibited, from coordinated disinformation to strategic manipulation.

Consider the adversarial scenario: an external actor seeks to shift public opinion within a target population but cannot directly contact those individuals due to detection risk, platform restrictions, or legal constraints. Instead, the actor must position agents within the surrounding social network, influencing targets only through authentic social connections. The spectral characterization developed above enables both offensive strategy optimization (identifying vulnerable targets and optimal resource allocation) and defensive vulnerability assessment (quantifying manipulation risk). We model this scenario formally as follows.

Entity $\mathcal{A}$ (the system, $\mathbf{x}_s \in \mathbb{R}^{N_s}$) represents the target population following natural DeGroot opinion dynamics. Entity $\mathcal{B}$ (the manipulator) seeks to shift $\mathcal{A}$'s opinions toward target $\mathbf{x}_s^*$ by deploying $N_z$ zealots— agents with fixed opinions $\mathbf{s}_Z \in \mathbb{R}^{N_z}$—positioned strategically at locations $\mathcal{P} \subset \{1,\ldots,N_b\}$ within the bath. Covertness requires zealots to be in the bath.

In standard DeGroot dynamics without zealots, the steady state is determined by initial conditions through a conservation law. For a primitive row-stochastic trust matrix $\mathbf{T}$, the Perron-Frobenius theorem guarantees a unique left eigenvector $\boldsymbol{\pi} > 0$ satisfying $\boldsymbol{\pi}^T\mathbf{T} = \boldsymbol{\pi}^T$ with $\boldsymbol{\pi}^T\mathbf{1} = 1$. While the right null vector $\mathbf{M}\mathbf{1} = \mathbf{0}$ determines the shape of the steady state (consensus), the left null vector $\boldsymbol{\pi}^T\mathbf{M} = \mathbf{0}^T$ determines which weighted combination of initial opinions is conserved: $\boldsymbol{\pi}^T\mathbf{x}(t) = \boldsymbol{\pi}^T\mathbf{x}(0)$ for all $t$. At steady state, $\mathbf{x}^* = c\,\mathbf{1}$ where $c = \boldsymbol{\pi}^T\mathbf{x}(0)$ is the influence-weighted average. Zealots fundamentally alter this picture: by maintaining fixed opinions for all time, they break the conservation law and pin the steady state independent of initial conditions.

To analyze this, we take the Laplace transform of Eq.~\eqref{eq:cont_GLE} with zealot forcing:
\begin{equation}
\mathsf{s}\tilde{\mathbf{x}}_s(\mathsf{s}) - \mathbf{x}_s(0) = \mathbf{M}_{ss}\tilde{\mathbf{x}}_s(\mathsf{s}) + \tilde{\mathbf{K}}(\mathsf{s})\tilde{\mathbf{x}}_s(\mathsf{s}) + \tilde{\boldsymbol{\eta}}(\mathsf{s}) + \tilde{\mathbf{f}}_z(\mathsf{s}),
\end{equation}
where the memory kernel transforms as
\begin{equation}
\tilde{\mathbf{K}}(\mathsf{s}) = \mathbf{M}_{sb}(\mathsf{s}\mathbf{I} - \mathbf{M}_{bb})^{-1}\mathbf{M}_{bs},
\end{equation}
and $\tilde{\mathbf{f}}_z(\mathsf{s})$ represents the zealot forcing. At steady state ($\mathsf{s} \to 0$), the Schur complement $\tilde{\mathbf{K}}(0) = -\mathbf{M}_{sb}\mathbf{M}_{bb}^{-1}\mathbf{M}_{bs}$ emerges, encoding all integrated bath-mediated pathways.

The crucial observation is that as $\mathsf{s} \to 0$, the terms involving initial conditions $\mathbf{x}_s(0)$ and $\mathbf{x}_b(0)$ (appearing in $\tilde{\boldsymbol{\eta}}(0)$) are dominated by the persistent zealot forcing $\tilde{\mathbf{f}}_z(0) \sim 1/\mathsf{s}$. This divergence reflects that zealots supply or absorb opinion ``current" indefinitely, effectively acting as reservoirs that overwrite the system's memory of its initial configuration.

When zealots occupy positions $\mathcal{P}$ within the bath with fixed opinions $\mathbf{s}_z$, the steady-state balance requires simultaneous satisfaction of system and bath equilibrium:
\begin{align}
\mathbf{0} &= \mathbf{M}_{ss}\mathbf{x}_s^* + \mathbf{M}_{sb}\mathbf{x}_b^* + \mathbf{M}_{sz}\mathbf{s}_z, \label{eq:inverse_system}\\
\mathbf{0} &= \mathbf{M}_{bs}\mathbf{x}_s^* + \mathbf{M}_{bb}\mathbf{x}_b^* + \mathbf{M}_{bz}\mathbf{s}_z, \label{eq:inverse_bath}
\end{align}
where $\mathbf{M}_{sz}$ and $\mathbf{M}_{bz}$ encode how zealot opinions couple to system and bath agents, respectively. These equations reflect opinion balance: at steady state, the net influence each agent receives from its neighbors (system, bath, and zealots) must sum to zero.

To analyze zealot influence, we temporarily set aside the system-bath 
partition and consider all agents without fixed opinions. Define 
\emph{free agents} $\mathbf{x}_F$ as the union of system and non-zealot 
bath agents—everyone whose opinion can evolve. The zealots $\mathbf{s}_Z$, 
with fixed opinions, act as boundary conditions. The free-agent dynamics 
are
\begin{equation}
\frac{d\mathbf{x}_F}{dt} = \mathbf{M}_{FF}\mathbf{x}_F + \mathbf{M}_{FZ}\mathbf{s}_Z,
\label{eq:free_dynamics}
\end{equation}
where $\mathbf{M}_{FF}$ governs interactions among free agents and $\mathbf{M}_{FZ}$ encodes zealot-to-free influence.

At steady state, each free agent's opinion satisfies the discrete Laplace equation with zealots as boundary conditions:
\begin{equation}
    x_i^* = \frac{1}{k_i}\sum_{j \sim i} x_j^*, \quad i \in F.
    \label{eq:harmonic}
\end{equation}
This admits a random-walk interpretation: $x_i^*$ equals the expected zealot opinion encountered by a walker starting at node $i$,
\begin{equation}
    x_i^* = \sum_{j \in Z} h_i^{(j)} s_{Z,j},
    \label{eq:hitting_sum}
\end{equation}
where $h_i^{(j)}$ is the probability of first hitting zealot $j$.  Defining the hitting probability matrix $\mathbf{H} \in \mathbb{R}^{N_F \times N_Z}$ with entries $H_{ij} = h_i^{(j)}$, the steady state becomes
\begin{equation}
    \mathbf{x}_F^* = \mathbf{H} \mathbf{s}_Z, \quad \text{where} \quad 
    \mathbf{H} = -\mathbf{M}_{FF}^{-1}\mathbf{M}_{FZ}.
    \label{eq:steady_hitting}
\end{equation}

Crucially, the steady state is generically a \emph{distribution}, not 
uniform consensus: agents closer (in random-walk terms) to high-opinion 
zealots equilibrate higher than those closer to low-opinion zealots, 
analogous to the temperature gradient between hot and cold boundaries 
in steady-state heat conduction.

For the inverse problem—finding zealot opinions that achieve a target distribution—we invert Eq.~\eqref{eq:steady_hitting}:
\begin{equation}
    \mathbf{s}_Z = \mathbf{H}^+ \mathbf{x}_F^*.
    \label{eq:inverse_hitting}
\end{equation}
Since $\mathbf{H}$ has rank at most $N_Z$, only distributions in its  column space are exactly achievable; the pseudoinverse yields the least-squares optimal zealot configuration for arbitrary targets.

\begin{figure*}[!t]
    \centering
    \includegraphics[width=0.99\linewidth]{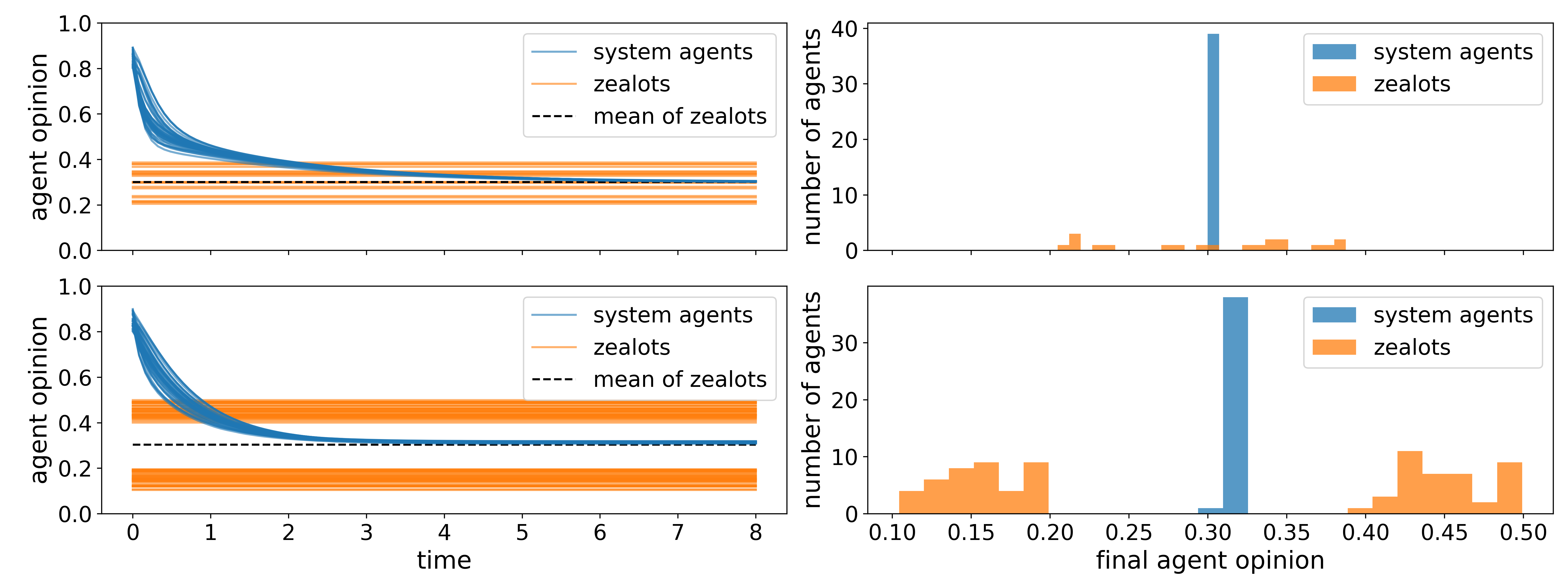}
    \caption{{\bf Zealot influence on system agents.} In the top row zealots opinions are sampled from a single uniform distribution, while in the bottom they are sampled from two disjoint uniform distributions. All system agents are sampled from a single uniform distribution. Opinion dynamics are shown in the left column, and steady-state opinion distributions are in the right column. Over time the influence of the zealots on system agents is seen by the tightening and shifting of the system agent opinion distribution. At steady-state the the system agent opinions are distributed around the mean zealot opinion.}
    \label{fig:op_dist_for_zealots}
\end{figure*}

Returning to the system-bath partition: since free agents comprise  both system and bath, Eq.~\eqref{eq:steady_hitting} determines both steady states simultaneously,
\begin{equation}
    \begin{pmatrix} 
        \mathbf{x}_s^* \\ 
        \mathbf{x}_b^* 
    \end{pmatrix} = 
    \begin{pmatrix} 
        \mathbf{H}_{sZ} \\ 
        \mathbf{H}_{bZ} 
    \end{pmatrix} 
    \mathbf{s}_Z,
    \label{eq:partition_hitting}
\end{equation}
where $\mathbf{H}_{sZ}$ and $\mathbf{H}_{bZ}$ are the rows of  $\mathbf{H}$ corresponding to system and bath agents, respectively.  The bath state is not assumed but \emph{predicted} by the harmonic structure. In covert operations where zealots are restricted to the bath, this framework reveals how indirect manipulation propagates: zealot influence diffuses through the bath before reaching the system, with the hitting probabilities $\mathbf{H}_{sZ}$ quantifying this indirect pathway.

Figure \ref{fig:op_dist_for_zealots} demonstrates the zealot's influence on system agents. In the top row we examine a scenario where zealot opinions are sampled from a single uniform distribution, and in the bottom row we examine when zealot opinions are sampled from two disjoint uniform distributions. System agents are initialized from a uniform distribution that is separate from the zealot's opinions in both scenarios. In the left column system (blue) and zealot (orange) opinions are shown through time, while in the right column the distribution of steady state opinions for system and zealot agents is shown. In both scenarios, the distribution of system agent opinions becomes more narrow through time, and the distributions shift towards the mean zealot opinion (black dashed line). At steady-state the system agent's opinions are tightly distributed around the mean of the zealot opinions.

We have developed a framework for transforming linear agent-based models from closed to open system descriptions through exact generalized Langevin equations. Environmental coupling manifests as memory kernels whose modal structure—timescales and coupling strengths—is determined by the spectrum of environmental interactions. This approach captures essential environmental information without explicit simulation of all environmental agents.

Applied to DeGroot opinion dynamics, the framework reveals how network topology shapes memory structure through two distinct mechanisms. Small-world rewiring enhances global information exchange, breaking eigenvalue degeneracy and driving the bath toward a single dominant relaxation mode. In contrast, fragmented environments with isolated subpopulations (echo chambers) sustain multiple persistent modes that cannot be collapsed into a single effective pathway. These structural signatures encode qualitatively different vulnerabilities to external manipulation.

For covert influence operations where adversaries position zealots exclusively within the environmental network, the steady-state system response follows from a hitting probability matrix that encodes random-walk distances to zealot positions. This structure reveals that zealots act as opinion reservoirs: system agents equilibrate to a distribution centered on the mean zealot opinion, with the spread determined by network geometry rather than initial conditions. The framework thus provides both a mechanistic understanding of indirect influence and a foundation for analyzing strategic vulnerabilities in networked systems.

Extensions to nonlinear dynamics through projection operator methods, time-varying networks, and optimization of zealot placement represent natural directions for future work.

This is document LA-UR-26-20919.

\bibliographystyle{apsrev4-2}
\bibliography{GLE}

\end{document}